\documentclass{article}
\usepackage{braket,bm,latexsym,amsmath,amssymb,amsfonts,fancyhdr,color,graphicx,multirow,slashed,cite}
\usepackage[a4paper,bottom=3cm,top=2.5cm,head=0mm,width=17cm,dvipdfm]{geometry}
\usepackage[usenames,dvipsnames,svgnames,table]{xcolor}
\usepackage[colorlinks=true,
            linkcolor=blue,
            urlcolor=blue,
            citecolor=green,          
						bookmarks=true,
						bookmarksnumbered=true,
						breaklinks=true,
						pdfpagemode=Fullscreen,
						pdfstartview=FitBH]{hyperref}
\usepackage[dotinlabels]{titletoc}
\usepackage{titlesec}
\usepackage{authblk,ulem}

\numberwithin{equation}{section}
\allowdisplaybreaks[4]

\titlelabel{\thetitle.\quad \hspace{-0.8em}}
\titlecontents{section}
              [1.5em]
              {\vspace{4mm} \large \bf}
              {\contentslabel{1em}}
              {\hspace*{-1em}}
              {\titlerule*[.5pc]{.}\contentspage}
\titlecontents{subsection}
              [3.5em]
              {\vspace{2mm}}
              {\contentslabel{1.8em}}
              {\hspace*{.3em}}
              {\titlerule*[.5pc]{.}\contentspage}
\titlecontents{subsubsection}
              [5.5em]
              {\vspace{2mm}}
              {\contentslabel{2.5em}}
              {\hspace*{.3em}}
              {\titlerule*[.5pc]{.}\contentspage}

\newcommand{\titledef}{High-Quality Axion Dark Matter without Isocurvature Problem} 

\hypersetup{ pdfauthor = {Shao-Feng Ge},
	     pdftitle = {\titledef}, 
	     pdfsubject = {}, 
             pdfkeywords = {}, 
	     pdfcreator = {LaTeX with hyperref package},
	     pdfproducer = {dvips + ps2pdf} }

\definecolor{gesfblack}{rgb}{0,0,0}

\definecolor{gesfblue}{rgb}{0.08,0.42,0.76}

\definecolor{gesfgreen}{rgb}{0,1,0}

\definecolor{gesfgrey}{rgb}{0.5,0.5,0.5}

\definecolor{gesflanse}{rgb}{0.00,0.50,0.50}

\definecolor{gesfpurple}{rgb}{0.47,0.19,0.42}

\definecolor{gesfred}{rgb}{1,0,0}

\definecolor{gesfwhite}{rgb}{1,1,1}

\definecolor{gesfyellow}{rgb}{0.7,0.4,0.3}

\newcommand{\gsec}[1]{{\hypersetup{linkcolor=red}Sec.\,\ref{#1}\hypersetup{linkcolor=blue}}}

\newcommand{\geqn}[1]{\hypersetup{linkcolor=blue}Eq.\,(\ref{#1})\hypersetup{linkcolor=blue}}
\newcommand{\gfig}[1]{{\hypersetup{linkcolor=violet}Fig.\,\ref{#1}\hypersetup{linkcolor=blue}}}

\definecolor{Orange}{cmyk}{0,0.61,0.87,0}
\definecolor{JungleGreen}{cmyk}{0.99,0,0.52,0}
\definecolor{OliveGreen}{cmyk}{0.64,0,0.95,0.40}
\definecolor{Brown}{cmyk}{0,0.81,1,0.60}
\definecolor{RoyalBlue}{cmyk}{0.71,0.53,0,0.12}
\definecolor{Gray}{cmyk}{0,0,0,0.40}
\definecolor{LightPink}{cmyk}{0.0,0.25,0,0}
\definecolor{LLightPink}{cmyk}{0.0,0.10,0,0}
\definecolor{LightBlue}{cmyk}{0.25,0,0,0}
\definecolor{LightGray}{cmyk}{0,0,0,0.2}

\graphicspath{{figs/}}

\usepackage{subfigure}
\newcommand{\bee}{\begin{equation}}
\newcommand{\ene}{\end{equation}}
\newcommand{\bea}{\begin{eqnarray}}
\newcommand{\ena}{\end{eqnarray}}

\begin{document}
\fontsize{12pt}{14pt}\selectfont

\title{
       \Large \bf \titledef} 

\author[1]{{\large Masahiro Kawasaki} \footnote{\href{mailto:??}{masahiro.kawasaki@ipmu.jp}}}
\author[1]{{\large Jie Sheng} \footnote{\href{mailto:jie.sheng@ipmu.jp}{jie.sheng@ipmu.jp}}}
\author[1,2]{{\large Tsutomu T. Yanagida} \footnote{\href{mailto:tsutomu.tyanagida@sjtu.edu.cn}{tsutomu.tyanagida@sjtu.edu.cn}}}

\affil[1]{Kavli IPMU (WPI), UTIAS, University of Tokyo, Kashiwa, 277-8583, Japan}
\affil[2]{Tsung-Dao Lee Institute \& School of Physics and Astronomy, Shanghai Jiao Tong University, Shanghai 200240, China}

\date{}

\maketitle

\vspace{-2mm}
\begin{abstract}
\fontsize{10pt}{12pt}\selectfont

Axion dark matter in high-scale inflation is subject to the isocurvature constraint, since quantum fluctuations of the axion field during inflation may exceed the current CMB bound. One conventional way to suppress these fluctuations is to assume that the Peccei–Quinn field has a large expectation value during inflation. However, this mechanism becomes ineffective when the axion domain wall number is large. In this work, we point out that a high-quality axion protected by a discrete gauge symmetry can naturally evade this problem. A Peccei–Quinn-violating but gauge-invariant operator induces a large effective axion mass during inflation, thereby suppressing the axion fluctuation. The same setup can address both the axion quality problem and the isocurvature problem, while leading to a prediction for the axion parameter space to be verified in future experiments.

\end{abstract}

\section{Introduction}

The particle nature of dark matter (DM) is one of the most important problems in modern physics~\cite{Planck:2018vyg,Cirelli:2024ssz}. The QCD axion is one of the most attractive candidates of DM, since it provides a simultaneous solution to the DM problem and the strong CP problem~\cite{Peccei:1977ur,Peccei:1977hh,Wilczek:1977pj,Weinberg:1977ma}. In QCD, the vacuum angle $\bar \theta$ is a physical CP-violating parameter and is generically expected to be of order unity. On the other hand, the current upper bound on the neutron electric dipole moment requires $|\bar \theta| \lesssim 10^{-10}$~\cite{Abel:2020pzs}, which poses a severe naturalness problem in the Standard Model. The Peccei--Quinn (PQ) mechanism solves this problem by promoting $\bar \theta$ to a dynamical field, the axion $a(x)$, associated with the spontaneous breaking of a global $U(1)_{\rm PQ}$ symmetry~\cite{Peccei:1977ur,Peccei:1977hh}. Non-perturbative QCD effects generate a potential for the axion, whose minimum dynamically cancels the effective vacuum angle.

Interestingly, the same axion can also account for the observed DM abundance through the misalignment mechanism~\cite{Preskill:1982cy,ABBOTT1983133,Dine:1982ah}. If the axion field $a(x)$ has an initial displacement from the minimum of its potential in the early universe, it starts coherent oscillations when the Hubble parameter becomes comparable to the axion mass, behaving as cold DM. The relic abundance $\Omega_a$ is approximately given by~\cite{Marsh:2015xka}
\begin{equation}
    \Omega_a h^2 \sim 0.12 \left( \frac{\theta_a}{0.8}\right)^2 \left( \frac{F_a}{10^{12}\,{\rm GeV}}\right)^{1.17},
    \label{axionRelic}
\end{equation}
up to anharmonic corrections, where $F_a$ is the axion decay constant and $\theta_a \equiv a_i/F_a$ is the initial misalignment angle. Thus, for $F_a \sim 10^{12}\,{\rm GeV}$ and an initial angle of order unity, the axion naturally explains the observed DM relic $\Omega_{\rm DM} h^2 \simeq 0.12$.

However, the cosmological history of the PQ symmetry breaking is highly constrained. If the $U(1)_{\rm PQ}$ symmetry is broken after inflation, cosmic strings are formed at the phase transition time~\cite{PhysRevLett.48.1867,Kawasaki:2018axionstrings}. When QCD effects become important, domain walls bounded by these strings also appear~\cite{PhysRevLett.48.1156}. In general, such domain walls are stable if the domain wall number satisfies $N_{\rm DW} > 1$, and their energy density would soon dominate the universe, leading to a serious cosmological problem~\cite{PhysRevLett.48.1156,Hiramatsu:2012gg}. This domain wall problem can be avoided only in a very special case $N_{\rm DW}=1$, where the string-wall system is unstable and decays. A simple and widely considered alternative is that the PQ symmetry is already broken during inflation and is never restored afterwards~\cite{Linde:1988}. In this case, the observable universe lies in a single PQ domain, and dangerous topological defects are inflated away.

This inflationary solution to the domain wall problem, however, leads to another difficulty. Many attractive models of inflation predict a high inflationary scale, corresponding to the Hubble parameter during inflation of order $H_{I}\sim 10^{13}\,{\rm GeV}$~\cite{Starobinsky:1980te, Bezrukov:2007ep, Kallosh:2013yoa}. Such high-scale inflation is especially interesting because it can be tested by future searches for primordial tensor modes~\cite{BICEPKeck:2021gln}. 
If the PQ symmetry is broken during inflation, the axion acquires quantum fluctuations, which generate axion isocurvature perturbations. For high-scale inflation and $F_a\sim 10^{12}\,{\rm GeV}$, these isocurvature fluctuations are typically much larger than the observational upper bound from the cosmic microwave background (CMB) if the axion is the dominant DM~\cite{Planck:2015sxf}. 
This is the so-called \textit{axion isocurvature problem}~\cite{Axenides:1983hj,PhysRevD.32.3178,Linde:1985yf,Linde:1990yj,PhysRevLett.66.5}. A well-known way to suppress the isocurvature perturbation is the Linde mechanism, in which the effective PQ scale during inflation is made much larger than its present value~\cite{Linde:1991km}. Nevertheless, this mechanism has potential subtleties. In particular, the post-inflationary dynamics of the PQ field may induce parametric resonance and restore the PQ symmetry~\cite{Kasuya:1998td,Kasuya:1999hy,Kawasaki:2013iha,Kawasaki:2018qwp}, which would recreate strings and domain walls. Moreover, we point out that if the domain wall number is large, the mechanism may also fail to suppress the isocurvature.

In this work, we propose a new mechanism to avoid the axion isocurvature problem in a high-quality axion model assisted by a discrete gauge symmetry $Z_N$~\cite{Krauss:1988zc,Kamionkowski:1992mf,Holman:1992us,Barr:1992qq,Babu:2002ic}. 
We show that, for sufficiently large $N$, as is usually required to ensure the high quality of the global $U(1)_{\rm PQ}$ symmetry~\cite{Dias:2002hz,Bhattiprolu:2021txa,Plakkot:2021xyx,Choi:2022jqy,Sheng:2025sou,Georis:2025kzv,Sheng:2026aro,Barger:2026flavor}, the dangerous isocurvature perturbations in this class of axion models can be sufficiently suppressed. The key observation is that the axion naturally obtains a large mass during inflation. In this way, the discrete gauge symmetry provides an economical mechanism that simultaneously solves the axion isocurvature problem and the axion quality problem.

The rest of this paper is organized as follows. In Sec.~2, we review the axion isocurvature problem. In Sec.~3, we introduce the Linde mechanism and discuss its potential challenges, including the possibility of symmetry restoration and its failure with large domain wall number. In Sec.~4, we present our new solution based on a high-quality axion assisted by a discrete gauge symmetry $Z_N$. The final section is devoted to our conclusions.

\section{Axion and Its Isocurvature Problem}

Let $\Phi$ be the complex scalar PQ field whose vacuum expectation value (vev) breaks the global PQ symmetry.  It is convenient to write it in polar form,
\begin{equation}
\Phi(x) = \frac{\rho}{\sqrt 2}\,e^{i\theta(x)}.
\label{eq:PQpolar}
\end{equation}
After spontaneous symmetry breaking, the radial mode $\rho$ gains vev $\langle \rho \rangle = v_{\rm PQ}$ and becomes heavy, while the angular mode $\theta (x)$ is the pseudo-Nambu Goldstone boson associated with the breaking of the PQ symmetry. This angular degree of freedom is the so-called axion field defined as $a(x) \equiv v_{\rm PQ} \theta (x)$. In such a way, the kinetic term for $\Phi$ near the vacuum, 
$\mathcal{L}_{\rm kin} = |\partial_\mu \Phi|^2 = \frac{1}{2} v_{\rm PQ}^2 (\partial_\mu \theta)^2$, can be expressed as $\frac{1}{2}(\partial_\mu a)^2$.
Because of the QCD anomaly, the axion couples to gluons as $\mathcal{L}\supset
\alpha_s\,
N_{\rm DW}\,\theta\, G^a_{\mu\nu}\tilde G^{a\mu\nu} / 8\pi
=
\alpha_s\, a G^a_{\mu\nu}\tilde G^{a\mu\nu}/ (8 \pi F_a)$.
It is then convenient to define the physical axion decay constant $F_a$ and axion angle by
\begin{equation}
F_a \equiv \frac{v_{\rm PQ}}{N_{\rm DW}}\,, 
\quad  \frac{a}{F_a} = N_{\rm DW}\,\theta.
\end{equation}

If the PQ symmetry breaks during the inflation and suppose the axion is light $m_a \ll H_I$, both the axion field $a$ and initial misalignment angle $\theta_a$ would acquire quantum fluctuations with amplitude~\cite{PhysRevLett.66.5},
\begin{equation}
 \delta a \sim \frac{H_I}{2\pi} \,, \quad \delta\theta_a = \frac{\delta a}{F_a} \sim \frac{H_I}{2\pi F_a}.
 \label{eq:deltaa}
\end{equation}
When the axion acquires a mass at the QCD scale and becomes a component of cold DM (CDM), fluctuations in the axion field induce axion density perturbations $\delta \rho_a$, which in turn contribute to the CDM isocurvature perturbation 
$S_{\mathrm{CDM}}$ as
\begin{equation}
S_{\mathrm{CDM}}
=
\frac{\delta \rho_{\mathrm{CDM}}}{\rho_{\mathrm{CDM}}}
=
\frac{\Omega_a}{\Omega_{\mathrm{CDM}}}
\frac{\delta \rho_a}{\rho_a}.
\end{equation}
The axion CDM can be produced through misalignment mechanism and the relic density $\Omega_a$ is determined only by the initial angle and decay constant as shown in \geqn{axionRelic}.
Similarly, the axion density is proportional to $\rho_a \propto \theta_a^2$ as the relic.
In this paper, we focus on the case in which the axion accounts for the entirety of DM with a benchmark decay constant of $F_a \simeq 10^{12}\,$GeV.
The condition $\Omega_a = \Omega_{\rm CDM}$ sets a relationship between $\theta_a$ and $F_a$ as, 
\begin{equation}
    \theta_a \simeq 0.82 \left(\frac{F_a}{10^{12}\,\text{GeV}} \right)^{-0.585}.
\label{theta_F}
\end{equation}
Therefore, one can re-express the isocurvature perturbation through the angle quantum fluctuation as,
\begin{equation}
    S_{\mathrm{CDM}} = \frac{\delta \rho_a}{\rho_a} = \frac{2\delta \theta_a}{\theta_a} = \frac{H_I}{\pi F_a \theta_a}.
\label{Scdm}
\end{equation}
The power spectrum of the CDM isocurvature perturbation is then,
\begin{equation}
\mathcal{P}_{\mathrm{iso}}(k)
\equiv 
\left\langle |S_{\mathrm{CDM}}|^2 \right\rangle
=
\left(
\frac{H_{\mathrm{I}}}{\pi F_a \theta_a}
\right)^2.
\end{equation}

On the other hand, the usual adiabatic curvature perturbation sourced by the inflation has a power spectrum of $\mathcal{P}_{\mathrm{adi}}(k_*) \simeq 2.1 \times 10^{-9}$ at the scale $k_* = 0.05\,\mathrm{Mpc}^{-1}$. The CMB observations have given stringent limit for the ratio between CDM isocurvature and adiabatic perturbations as~\cite{Planck:2015sxf},
\begin{equation}
\beta_{\mathrm{iso}}
\equiv
\frac{\mathcal{P}_{\mathrm{iso}}(k_*)}
{\mathcal{P}_{\mathrm{iso}}(k_*)+\mathcal{P}_{\mathrm{adi}}(k_*)}
< 0.038  \quad 
\Rightarrow \quad 
\mathcal{P}_{\mathrm{iso}}(k_*) < 8.3 \times 10^{-11},
\label{CMB_limit}
\end{equation}
Substituting \geqn{Scdm} and \geqn{theta_F}, the \geqn{CMB_limit} gives a constraint for the inflation Hubble,
\begin{equation}
H_I
\lesssim
2.4\times 10^7\,\mathrm{GeV}
\left(
\frac{F_a}{10^{12}\,\mathrm{GeV}}
\right)^{0.415},
\end{equation}
This constraint implies that the axion DM scenario in which the PQ symmetry is broken during inflation is in tension with high-scale inflation, since it would generate isocurvature perturbations that are too large to be consistent with CMB observations. This is the axion isocurvature problem.

\section{The Linde Mechanism in Perspective}

In this section, we introduce the Linde mechanism~\cite{Linde:1991km}, which can suppress axion isocurvature perturbations in high-scale inflation models. However, this mechanism may suffer from the problem of PQ symmetry restoration caused by parametric resonance, and may also face difficulties when the domain wall number is large.

\subsection{Linde mechanism}
The origin of the excessively large axion isocurvature perturbation is that the ratio of the Hubble scale during inflation to the axion expectation value associated with the angular fluctuation \geqn{eq:deltaa} is too large. As first pointed out by Linde, if the PQ field has a large initial vacuum expectation value, of order the Planck scale, $|\Phi_i| \sim M_{\rm p}$ ($M_{\rm p} \simeq 2.4 \times 10^{18}\,$GeV), during inflation and subsequently oscillates down to a smaller value after inflation, the axion fluctuation can be significantly suppressed, as follows:
\begin{equation}
\delta \theta_a = \frac{N_{\rm DW} H_{I}}{2\pi |\Phi_i|}.
\label{theta_a}
\end{equation}
Using the limit $\mathcal{P}_{\mathrm{iso}} = (2 \delta \theta_a / \theta_a)^2 < 8.3 \times 10^{-11}$ and the relationship \geqn{theta_F} between $\theta_a$ and $F_a$, one can obtain the constraint on Hubble parameter during inflation,
\begin{equation}
H_I \lesssim 5.7 \times 10^{13}\,\mathrm{GeV}\,
\frac{1}{N_{\rm DW}}
\left( \frac{F_a}{10^{12}\,\mathrm{GeV}} \right)^{-0.585}
\label{H_fa}
\end{equation}
For $|\Phi_i| = M_{\rm p}$. Due to the suppression of the isocurvature perturbation, the allowed Hubble scale during inflation is greatly relaxed, making it consistent with high-scale inflation scenarios.

The inflation scale can be inferred from CMB observations of primordial tensor perturbations, i.e. inflationary gravitational waves, since the tensor-to-scalar ratio $r$ is related to the Hubble parameter during inflation as~\cite{Baumann:2009ds}
\begin{equation}
r
=
1.6\times 10^{-3}
\left(
\frac{H_{I}}{10^{13}\,\mathrm{GeV}}
\right)^2.
\label{r_H}
\end{equation}
The current CMB bound reaches $r < 0.036$~\cite{BICEPKeck:2021gln} implies an upper limit of $H_I < 4.7 \times 10^{13}\,$GeV. 
This means that high-scale inflation predicting a tensor-to-scalar ratio of order $r\sim 10^{-3}$
could be probed in the near future. Since \geqn{H_fa} shows that the Hubble scale is inversely proportional to the axion decay constant, one obtains the following relation between $r$ and $F_a$ as,
\begin{equation}
r \lesssim 0.001\,
\frac{1}{N_{\rm DW}^2}
\left( \frac{F_a}{2.72\times 10^{13}\,\mathrm{GeV}} \right)^{-1.17}.
\label{Tmode}
\end{equation}
Therefore, in the high-scale inflation scenario of interest, under the requirement that axions account for the entire DM abundance, the axion decay constant $F_a$ is roughly in the range $10^{11} \sim 10^{13}\,$GeV.
Therefore, Linde mechanism can make the axion consistent with the sensitivity of future observation.

\subsection{Symmetry restoration by parametric resonance and a possible solution}
\label{sec:parameter}

However, the above mechanism may suffer from the serious problem of PQ symmetry restoration through parametric resonance. The simplest standard potential generates the vev through a mass term and the quartic term of the $\Phi$ field:
\begin{equation}
V(\Phi) = -m^2 |\Phi|^2 + \frac{1}{2}\lambda |\Phi|^4 + c_H H^2 |\Phi|^2 .
\end{equation}
Here, $m$ is the PQ field mass and $\lambda$ is the coupling constant. Since the PQ field is accompanied by the inflaton, their interaction should also be taken into account. This interaction induces an additional Hubble-induced mass term with a negative coefficient $c_H$. During inflation, the PQ field acquires a large field value $|\Phi_i| = \sqrt{-c_H/\lambda} H_I$, which can be as large as the Planck scale if $\lambda$ is sufficiently small, thereby realizing naturally the Linde mechanism.

However, as the Hubble parameter decreases rapidly after inflation, the quartic term drives oscillations of the PQ field. At the same time, fluctuations of the PQ field can be exponentially amplified by parametric resonance. As a result, this potential receives a correction of the form
$\Delta V \simeq \lambda \langle |\delta \Phi|^2 \rangle |\Phi|^2$.
If the fluctuation grows up to $\langle |\delta \Phi|^2 \rangle \sim v^2$,
the $U(1)_{\rm PQ}$ symmetry is non-thermally restored~\cite{Kasuya:1998td,Kasuya:1999hy,Kawasaki:2013iha,Kawasaki:2018qwp}. Once this happens, cosmic strings and domain walls are formed, leading to the domain wall problem.

To see the instability explicitly, one can solve the equation of motion for the $\Phi$ field.
First, we write the complex field as $\Phi = X+iY$
and choose the initial homogeneous direction along the real axis, $\bar X_i = |\Phi|_i,
\bar Y_i = 0.$
Both the real and imaginary parts can be decomposed into homogeneous pieces plus fluctuations:
\begin{equation}
X=\bar X+\delta X,
\qquad
Y=\bar Y+\delta Y,
\end{equation}
By introducing the dimensionless time variable $z$ and the rescaled fluctuations, 
\begin{equation}
    z = c \sqrt{\lambda}|\Phi_i| \int_{t_i}^t \frac{dt}{a}, \quad \delta x_k \equiv a\,\delta X_k,
\quad
\delta y_k \equiv a\,\delta Y_k.
\end{equation}
the linearized equation of motion for the fluctuations of the PQ field can be cast into the standard Mathieu form:
\begin{equation}
\frac{d^2\delta x_k}{dz^2} + [A_x+2q_x\cos 2z]\,\delta x_k \simeq 0,
\label{Meq_x}
\end{equation}
\begin{equation}
\frac{d^2\delta y_k}{dz^2} + [A_y+2q_y\cos 2z]\,\delta y_k \simeq 0,
\label{Meq_y}
\end{equation}
with
\begin{equation}
A_x = \frac{k^2-a^2\lambda v^2}{c^2\lambda|\Phi|_i^2}+2q_x,
\quad
A_y = \frac{k^2-a^2\lambda v^2}{c^2\lambda|\Phi|_i^2}+2q_y,\quad
q_x=\frac{3}{4c^2},
\quad
q_y=\frac{1}{4c^2}.
\label{paras}
\end{equation}
Here, $c \simeq 0.84$ is a constant and hence $q_x = 1.04$, $q_y = 0.35$. 

Although the coefficient of the linear term in the Mathieu equation \geqn{Meq_x} or \geqn{Meq_y} is always positive, the equation still admits exponentially growing unstable solutions of the form
$\propto e^{\mu_k z}$,
when the parameters $(A,q)$ fall into the instability bands. This is the origin of parametric resonance. The instability bands appear around
$A = n^2,\, n = 1,2,\dots$
From \geqn{paras}, we obtain $A_x \gtrsim 2.1$ and $A_y \gtrsim 0.7$.
Therefore, the parametric resonance occurs in the second instability band for $\delta x_k$, while it occurs in the first instability band for $\delta y_k$.
Since the resonance is stronger in the first instability band ($1 - q < A_y < 1 + q$), we focus on  $\delta y_k$.
The fluctuation of the imaginary component $\delta y$ has an unstable solution and grows exponentially as,
\begin{equation}
    \delta y_k \propto e^{\mu_k z}
\end{equation}
with $\mu = 1/8 c^2$ is a constant of $\mathcal{O}(1)$. The instability develops when $\mu_k z \gg 1$. Lattice simulations show that the condition for this is $z \gtrsim 100$~\cite{Kasuya:1998td,Kasuya:1999hy,Kawasaki:2013iha,Kawasaki:2018qwp}.

The $\sqrt{\lambda}|\Phi|^4$ potential can be written as an effective mass term $m^2_{\rm eff} |\Phi|^2$ with $m_{\rm eff} \equiv \sqrt{\lambda} |\Phi|$. At the same time, the scale factor dependence of $|\Phi|$ is $|\Phi| \propto a^{-1}$ for the fourth-power potential. Therefore, the dimensionless time variable $z$ grows as $z \simeq \int m_{\rm eff} dt \simeq m_{\rm eff} H^{-1}$ in one Hubble time. After inflation ends, the inflaton begins to oscillate, and the Universe enters a matter-dominated era. In this regime, the Hubble parameter scales as $H \simeq \sqrt{\lambda} |\Phi_i| (a/a_i)^{-3/2} = \sqrt{\lambda} |\Phi_i| (|\Phi|/|\Phi_i|)^{3/2}$. Combining the two expressions, the condition for the PQ field to avoid parametric resonance is:
\begin{equation}
    z \simeq \frac{m_{\rm eff}}{H} 
    = \left( \frac{|\Phi_i|}{v_{\rm PQ}} \right)^{1/2} \lesssim 10^2 \quad \Rightarrow \quad \frac{|\Phi_i|}{v_{\rm PQ}} \lesssim 10^4.
    \label{condition1}
\end{equation}
For \(\Phi_i = M_{\rm P}\), the condition requires the eventual PQ vev $v_{\rm PQ}$ to satisfy $v_{\rm PQ} \equiv F_a N_{\rm DW} \gtrsim 10^{14}~\mathrm{GeV}$.
However, this is in conflict with our assumption of high-scale inflation \geqn{H_fa}. Therefore, the Linde mechanism with the simplest $\Phi^4$ potential does not work.


This problem can be avoided by introducing a higher-dimensional term in the PQ field potential, such as $\Phi^6$ term, together with a small mass $m$ for the PQ field~\cite{Kawasaki:2018qwp}:
\begin{equation}
V(\Phi) = -m^2 |\Phi|^2 + \frac{1}{2}\lambda |\Phi|^4 + \frac{1}{3}\eta \frac{|\Phi|^6}{M_{\rm p}^2} + c_H H^2 |\Phi|^2 \, .
\end{equation}
Again, to realize Linde mechanism $|\Phi_i| \simeq M_{\rm p}$, the coupling $\eta$ should be tiny as 
$\eta \simeq H_I^2 / M_{\rm p}^2 \simeq 10^{-10}$.
After inflation, PQ field starts to oscillate. When the amplitude $|\Phi|$ is large, the $|\Phi|^6$ term still dominates the potential. As the amplitude of $\Phi$ decreases, the quartic term becomes dominant when:
\begin{equation}
    \frac{1}{2}\lambda |\Phi_4|^4 \simeq \frac{1}{3}\eta \frac{|\Phi_4|^6}{M_{\rm p}^2} 
    \quad \Rightarrow \quad
    |\Phi_4| \simeq \sqrt{\frac{3\lambda}{2\eta}} M_{\rm p}.
\end{equation}
Once the field value of $\Phi$ becomes smaller than this threshold, the oscillation is driven by the quartic term, and parametric resonance can still occur. However, in this case, the onset of the oscillation is no longer at the Planck scale, but at $\Phi_i = \Phi_4$. According to \geqn{condition1}, the condition for avoiding parametric resonance is then given by:
\begin{equation}
    v_{\rm PQ} \gtrsim 10^{-4} |\Phi_4| \simeq 
    10^{-4} \sqrt{\frac{\lambda}{\eta}} M_{\rm p}
\end{equation}
In this case, if the coupling $\lambda$ is sufficiently small as $\lambda \lesssim 10^{-14}$, the PQ vev of interest, $v_{\rm PQ} \sim F_a \sim 10^{12}~\mathrm{GeV}$, can also satisfy the above condition. A small coupling also implies a correspondingly small mass for the PQ field, $m \sim \sqrt{\lambda} v_{\rm PQ} < 100\,$TeV.


\subsection{Challenges with large domain wall number}

In addition to the issue of PQ symmetry restoration, the Linde mechanism may also become ineffective in certain axion models with a large domain wall number.
The CMB constraint on isocurvature perturbations can be directly translated into an upper bound on the quantum fluctuation of the axion misalignment angle:
\begin{equation}
   \mathcal{P}_{\mathrm{iso}} 
   = \left( \frac{2 \delta \theta_a}{\theta_a}\right)^2 
   < 8.3 \times 10^{-11}
   \quad \Rightarrow \quad  
   \delta \theta_a 
   < 4.6\times 10^{-6} \times
   \left( \frac{\theta_a}{1}\right).
\label{angle_cons}
\end{equation}

Let us assume that the Planck scale is the largest physical scale in the theory. Then the maximal possible initial value of the PQ field during inflation is bounded by $\Phi_i \lesssim M_{\rm P}$.
Even in this most optimistic case, however, the Linde mechanism can be ineffective when the domain wall number is large. This is because the axion decay constant and then the angle fluctuation during inflation is determined not only by the PQ-field value but also by the domain wall number as \geqn{theta_a}. 
Comparing this with the CMB isocurvature constraint in \geqn{angle_cons}, we obtain the upper bound on the domain wall number for the Linde mechanism to work:
\begin{equation}
    N_{\rm DW} < 7.0 \times \left( \frac{\theta_a}{1}\right) \left( \frac{10^{13}\,{\rm GeV}}{H_I}\right)\left( 
    \frac{|\Phi_i|}{M_{\rm p}}\right). 
\end{equation}
Therefore, for high-scale inflation with $H_{\rm inf}\sim 10^{13}\,{\rm GeV}$ and an initial misalignment angle of order unity, the Linde mechanism can only work for models with a domain wall number smaller than $7$. If we take the maximum value of $\theta_a = \pi$, the upper limit of domain wall number is $N_{\rm DW} \simeq 21$.

The domain wall number $N_{\rm DW} \simeq 15$ is one of prediction in the framework of standard model which preserves the sufficient high quality of the axion to solve the strong CP problem \cite{Sheng:2025sou}. Furthermore, it is  the anomaly free discrete symmetry in the $E_7$ non-linear unification \cite{Kugo:1984sw}.

\section{High-Quality Axion with Discrete Symmetry as a New Solution}

In this section, we first review the axion quality problem, and then introduce a discrete gauge symmetry as a well-motivated mechanism to protect the quality of the PQ symmetry. We point out that this mechanism can also provide a natural solution to the axion isocurvature problem.

\subsection{High-quality problem and discrete gauge symmetry}

The PQ mechanism assumes that the low-energy theory possesses an approximate global $U(1)_{\rm PQ}$ symmetry.  After it is spontaneously broken, the axion dynamically relaxes
the effective QCD vacuum angle to zero.  The point of the axion quality problem is that this solution
is extremely sensitive to any additional explicit breaking of $U(1)_{\rm PQ}$.
It is widely believed that Planck-scale quantum gravity physics, like black holes and wormholes,  
would violate explicitly all global symmetries including the PQ symmetry~\cite{Giddings:1988cx,Coleman:1988tj,Gilbert:1989nq,Hawking:1975vcx}.
At energies small compared to the Planck mass $M_{\text{pl}}$, these symmetry-violating effects should be described by higher-dimensional potential operators of the $U(1)_{PQ}$ field $\Phi$ as
\begin{equation}
    V_{\rm p} (\Phi)
    =
    -\frac{g_n}{M_{\rm p}^{\,n-4}}
    \Phi^n
    +
    {\rm h.c.},
    \label{eq:general-planck-operator}
\end{equation}
with integer $n = 5, 6, \cdots$ except for the usual PQ field potential, $V(\Phi) = -m^2 |\Phi|^2 + \frac{1}{2}\lambda |\Phi|^4$. 
After the spontaneous PQ symmetry breaking by the $\Phi$ condensation, $\Phi=\frac{v_{\rm PQ}}{\sqrt{2}} e^{i \frac{a}{v_{\rm PQ}}}$, the Planck scale physics would induce an extra 
potential $V_g^a$ for the axion as~\cite{Kamionkowski:1992mf,Holman:1992us}
\begin{equation}
    V^a_{\rm p}(a)
    =
    2 |g_n| \frac{(v_{\rm PQ}/\sqrt{2})^n}{M_{\rm p}^{n-4}}
    \left[
        1-\cos\left(
            n\frac{a}{v_{\rm PQ}}+\delta
        \right)
    \right],
    \label{eq:gravity-induced-potential}
\end{equation}
with the $\delta$ being the phase of coupling $g$. The gravity-induced
axion mass $m_a^{\rm p}$,
\begin{equation}
    \left(m_a^{\rm p}\right)^2
    =
    n^2 |g_n|\,M_{\rm p}^2
    \left(
        \frac{v_{\rm PQ}}{\sqrt{2} M_{\rm p}}
    \right)^{n-2}.
    \label{eq:gravity-induced-mass}
\end{equation}

The gravity-induced potential \geqn{eq:gravity-induced-potential} certainly shifts the QCD axion potential. To solve the strong CP problem, the value of the effective phase, $\theta_{\text{eff}} \equiv \bar \theta + N_{\rm DW}\braket{a}/v_{\rm PQ}$, must be smaller than $\sim 10^{-10}$~\cite{Abel:2020pzs}. In order for the gravity-induced potential not to violate this constraint, it implies that the ratio between the gravity-induced axion mass and the QCD axion mass, $R \equiv (m^{\rm p}_a/m_a)^2$, should also be less than $10^{-10}$ as,
\begin{equation}
    n^2 |g_n|\,M_{\rm p}^2
    \left(
        \frac{F_a N_{\rm DW}}{\sqrt{2}M_{\rm p}}
    \right)^{n-2} 
\lesssim
    10^{-10} \times \left[5.7\,\mu{\rm eV} \times \left(\frac{10^{12}\,{\rm GeV}}{F_a} \right)\right]^2.
\label{highQ_req}
\end{equation}
Here, we apply the QCD induced axion mass 
$m_a = [\sqrt{m_u m_d}/ (m_u+m_d)] \times m_\pi f_\pi / F_a = 5.7\,\mu\text{eV} \times (10^{12}\,\text{GeV}/F_a)$,
where up-quark mass $m_u = 2.16\,$MeV, down quark mass $m_d = 4.67\,$MeV, pion mass $m_\pi \simeq 140\,\mathrm{MeV}$, and pion decay constant $f_\pi = 92\,\mathrm{MeV}$.

If no discrete gauge symmetry is imposed, the leading
PQ-violating operator is generically expected to appear already at dimension
five with $n = 5$,
which carries the smallest possible Planck suppression among nonrenormalizable
operators. This is the most dangerous case, since the gravity-induced axion
potential is then only suppressed by one power of $M_{\rm p}$.
Ignoring order-one factors, including the domain-wall number, and taking
$v_{\rm PQ}\simeq F_a$, the axion quality condition requires approximately
$F_a \lesssim 10\,{\rm GeV}$.
This bound is clearly incompatible with the axion window of interest, $F_a=10^{12}\,{\rm GeV}$. It implies that 
the dimension-five Planck-induced operator overwhelms the QCD axion potential and shifts the axion minimum by an unacceptably large
amount, thus spoiling the axion solution to the
strong CP problem.

This motivates imposing an exact discrete gauge symmetry $Z_N$, whose role is to
forbid the dangerous low-dimensional PQ-violating operators and postpone the
first allowed breaking term to sufficiently high dimension.
In such a case, only the higher order dimension operator with $n = kN$ with $k = 1, 2 \cdots$ remains, and the most dangerous one is,
\begin{equation}
    V_{\rm p}^N (\Phi)
    =
    -\frac{g_N}{M_{\rm p}^{\,N-4}}
    \Phi^N
    +
    {\rm h.c.}.
    \label{eq:general-planck-operator}
\end{equation}
Here, the integer $N$ appearing in $Z_N$ is, by definition, identified with the above anomaly coefficient, namely
\begin{equation}
N \equiv N_{\rm DW}.
\end{equation}
The axion domain wall number is determined by the QCD anomaly coefficient of the PQ current. 
With the axion normalized such that $a/F_a$ has period $2\pi$, the color anomaly coefficient is defined as
$N_{\rm DW}
\equiv 
\left| \sum_i \left(\chi_{iL}-\chi_{iR}\right) \right|$ in the fundamental representation of $SU(3)_c$,
where $\chi_{iL}$ and $\chi_{iR}$ are the PQ charges of the left- and right-handed components of the colored fermion $\psi_i$. The PQ current therefore has the QCD anomaly
$\partial_\mu J^\mu_{\rm PQ}
=
N_{\rm DW}\,
\frac{g_s^2}{32\pi^2}
G^a_{\mu\nu}\widetilde{G}^{a\mu\nu}$.
Equivalently, the axion appears in the effective QCD theta angle as $\bar \theta_{\rm eff}=\bar \theta+ N_{\rm DW}\langle a\rangle/v_{\rm PQ}$. Since the QCD vacuum energy is periodic in $\theta_{\rm eff}$ with period $2\pi$, there are $N_{\rm DW}$ inequivalent minima within one fundamental axion period. This is the origin of the axion domain wall number.
This $Z_N$ is precisely the anomaly-induced discrete subgroup associated with the PQ color anomaly.

When $N$ is sufficiently large, the leading PQ-violating operator allowed by the discrete symmetry is suppressed by a sufficiently high power of the Planck scale. As a result, the axion mass induced by Planck-scale physics becomes negligibly small and can satisfy the axion-quality condition in \geqn{highQ_req}. For the benchmark choice $g_N = 10^{-3}$ and $F_a = 10^{12}\,\mathrm{GeV}$, we find that the minimum value required to satisfy the quality bound is $N_{\min} = 16$. Therefore, imposing a discrete gauge symmetry with sufficiently large order $N$ is one of the standard approaches to addressing the axion quality problem.

This also shows that the Linde mechanism, as a solution to the isocurvature problem, is intrinsically incompatible with a high-quality axion with a large domain wall number. Therefore, an alternative mechanism is required.

\subsection{Isocurvature suppressed by a large axion mass during inflation}

We now point out that the same discrete gauge symmetry which protects the quality of the
PQ symmetry can also provide a simple solution to the axion isocurvature problem. The key
observation is that the leading operator allowed by the discrete symmetry but violating the
continuous $U(1)_{\rm PQ}$ symmetry gives the axion a large effective mass during inflation.
This mass suppresses the axion fluctuation before it becomes an observable isocurvature
perturbation.

Including the leading $Z_N$-invariant but PQ-violating operator, the PQ potential can be
written as
\begin{equation}
V(\Phi)
=
-m^2 |\Phi|^2
+
\frac{1}{2}\lambda |\Phi|^4 - H_I^2 |\Phi|^2
+
\left(
g_N \frac{\Phi^N}{M_{\rm p}^{N-4}}
+ {\rm h.c.}
\right) .
\label{eq:PQ_potential_discrete}
\end{equation}
Here we still take the cutoff scale to be the reduced Planck
scale. 
Here the coefficient $g_N$ cannot be taken to be of order unity. Since the PQ field is assumed to take a large value during inflation, $|\Phi_i|\sim M_p$, the $Z_N$-invariant but PQ-violating operator contributes to the vacuum energy by
$
|V_N| \sim 2 |g_N| |\Phi_i|^N/M_p^{N-4}
\sim 2 |g_N| M_p^4$.
This contribution should not dominate over the inflaton energy density, $V_{\rm inf} \simeq 3 H_I^2 M_p^2$.
Therefore, up to order-one factors, we require
\begin{equation}
|g_N| < \frac{3 H_I^2}{M_p^2}
\simeq 5.3 \times 10^{-11}
\left(\frac{H_I}{10^{13}\,{\rm GeV}}\right)^2 .
\end{equation}

Writing $\Phi = |\Phi| e^{i\theta}$, 
the angular part of the potential is given by
\begin{equation}
V_{\rm ang}(\theta)
\simeq
-2 |g_N| \frac{|\Phi|^N}{M_{\rm p}^{N-4}}
\cos \left[N(\theta-\theta_0)\right],
\label{eq:angular_potential_inflation}
\end{equation}
where $\theta_0$ is determined by the complex phase of $g_N$. 
This expression shows explicitly
that the PQ-breaking operator lifts the otherwise flat axion direction. In other words, the
angular direction is no longer a massless Nambu--Goldstone direction during inflation. 
When \geqn{eq:angular_potential_inflation} is extended back to the full radial field space, certain angular directions, for example those with $\cos[N(\theta-\theta_0)]>0$, can make the radial potential unstable at large values of $|\Phi|$. However, in a complete UV theory, additional higher-dimensional operators consistent with the imposed symmetry are generically allowed and can stabilize the radial direction. In the present discussion, we focus only on the minimal Planck-suppressed operator most relevant to the axion quality problem and to the suppression of isocurvature perturbations.


During inflation, the Linde mechanism naturally allows the radial field to take a large value, $|\Phi_i| \sim M_p$.
At this large field value, the PQ-breaking operator in Eq.~\eqref{eq:angular_potential_inflation}
induces a sizable curvature for the angular direction. Expanding this potential around one of its minima,
$\theta = \theta_{\rm min} + \delta\theta $,
we obtain the quadratic part of the angular potential as
\begin{equation}
    V_{\rm ang}
    \supset
    |g_N| N^2
    \frac{|\Phi_i|^N}{M_{\rm p}^{N-4}}
    (\delta\theta)^2 .
\end{equation}
However, $\theta$ itself is not the canonically normalized field. The kinetic term of the PQ
field contains $|\partial_\mu \Phi|^2
    \supset
    |\Phi_i|^2 (\partial_\mu \theta)^2 $,
where the radial field has been fixed at $|\Phi_i|$ during inflation. Therefore the canonically
normalized angular fluctuation is, $\delta a_{\rm inf}
    \equiv
    |\Phi_i| \delta\theta $
Substituting this relation into the quadratic angular potential gives
\begin{equation}
    V_{\rm ang}
    \supset
    |g_N| N^2
    \frac{|\Phi_i|^{N-2}}{M_{\rm p}^{N-4}}
    (\delta a_{\rm inf})^2 .
\end{equation}
Comparing this with the standard mass term,
we obtain the effective mass of axion during inflation to be,
\begin{equation}
    m^2_{a,{\rm eff}}
    \simeq
    N^2 |g_N|
    \frac{|\Phi_i|^{N-2}}{M_{\rm p}^{N-4}} .
\end{equation}
This should be contrasted with the usual QCD axion mass, which is negligible during inflation.
Controlled by the large inflationary field
value $|\Phi_i|$, this mass is no longer negligible.

The evolution of the angular fluctuation during inflation is then modified. For a massive
angular mode, the superhorizon fluctuation is exponentially damped as~\cite{Anisimov:2004hr,Jeong:2013xta,Higaki:2014ooa,Nakayama:2015pba,Kearney:2016vqw}
\begin{equation}
\delta\theta_a
\simeq
\frac{N_{\rm DW} H_I}{2\pi |\Phi_i|}
\exp\left[
-\frac{m^2_{a,{\rm eff}}}{3H_I^2}N_*
\right].
\label{eq:thetaPQ_suppression}
\end{equation}
where $N_* = 50 \sim 60$ is the number of $e$-folds between the horizon exit of the CMB scale and the end
of inflation. For illustration, we take $N_* = 60$. In the Linde mechanism, the PQ field takes a large initial value during inflation, $|\Phi_i| \lesssim M_{\rm p}$.
The corresponding effective axion mass can approach the order of inflationary Hubble scale, $m_{a,{\rm eff}} \lesssim H_I$.
As a result, the angular fluctuation is exponentially suppressed during inflation, and the resulting axion isocurvature perturbation becomes negligibly small. This provides the physical picture in which a discrete gauge symmetry can address both quality and isocurvature problems.
 
Numerically, in order to identify the viable parameter space, we need to impose both the axion quality bound and the isocurvature constraint simultaneously. These requirements can be summarized by the following two inequalities:
\begin{align}
    N^2 |g_N|\,M_{\rm p}^2
    \left(
        \frac{F_a N}{\sqrt{2} M_{\rm p}}
    \right)^{N-2} 
&\lesssim
    10^{-10} \times \left[5.7\,\mu{\rm eV} \times \left(\frac{10^{12}\,{\rm GeV}}{F_a} \right)\right]^2 \,, 
    \label{eq1} \\
\frac{N H_I}{2 \pi |\Phi_i|}
\exp\left[
-\frac{m^2_{a,{\rm eff}}}{3H_I^2}N_*
\right] &\lesssim 4.6 \times 10^{-6} \times \theta_a.
\label{eq2}
\end{align}
Here we have used the relation $N_{\rm DW} = N$, and can further employ the following relations:
\begin{equation}
    |\Phi_i| = 10^4 F_a N\,, \quad 
    m^2_{a,{\rm eff}}
    \simeq
    N^2 |g_N|
    \frac{|\Phi_i|^{N-2}}{M_{\rm p}^{N-4}} \,, \quad 
    \theta_a = 0.8 \left( \frac{F_a}{10^{12}\,{\rm GeV}} \right)^{-0.585}.
\end{equation}
The first condition here, $|\Phi_i| = 10^4 F_a N$, originates from the constraint on parametric resonance discussed above. On the one hand, a larger initial field value $|\Phi_i|$ is preferred in order to obtain a large effective axion mass during inflation. On the other hand, the axion quality condition favors a smaller value of $F_a N_{\rm DW}$. We therefore take the ratio between these two scales to be the largest value consistent with avoiding PQ symmetry restoration, $10^4$.
The relation between $\theta_i$ and $F_a$ is fixed by the condition that the axion constitutes the observed DM abundance. In addition, we adopt the following benchmark values for the remaining parameters as mentioned before:
\begin{equation}
    H_I = 10^{13}\,{\rm GeV}\,, \quad  
    |g_N| = 10^{-11}\,, \quad N_* = 60\,,\quad |\Phi_i| < M_{\rm p}.
\end{equation}
In such a way, the two inequalities are reduced to constraints on only two variables, $N$ and $F_a$. The viable parameter space can then be obtained by solving these inequalities simultaneously.

\begin{figure}[t]
\centering
\includegraphics[width=0.8\textwidth]{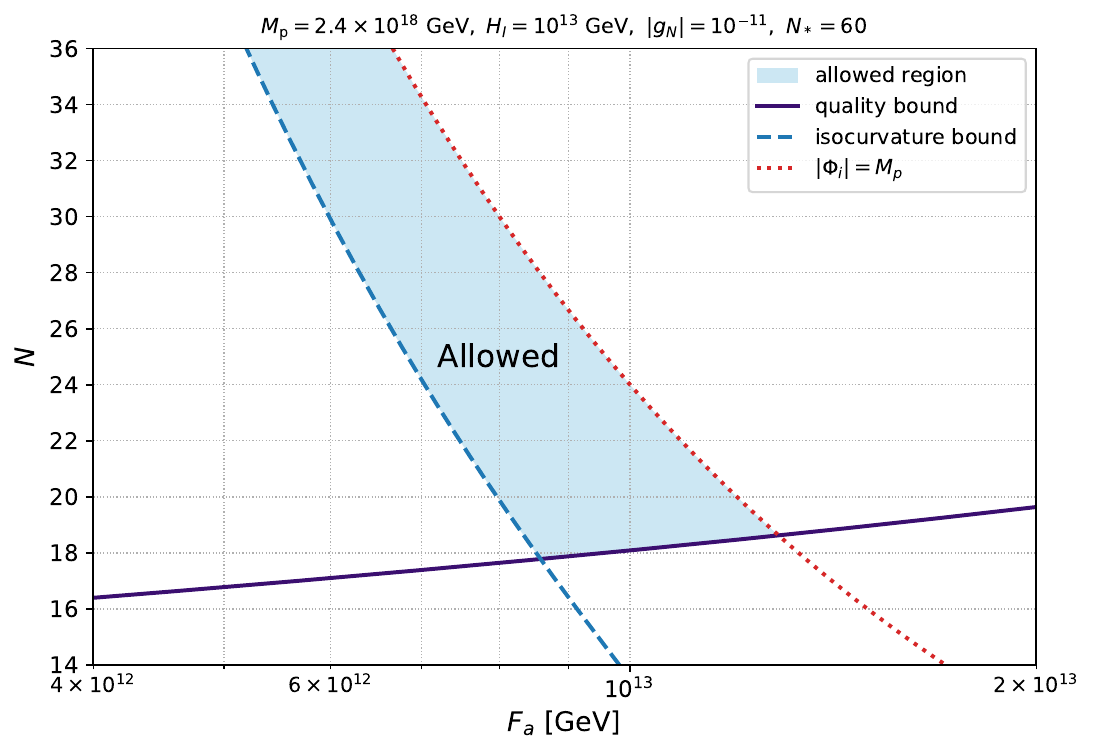}
\caption{
Parameter space in the axion decay constant $F_a$ and the discrete symmetry number $N$ satisfying both the axion high quality and isocurvature constraints. We assume $|g_N| = 10^{-11}$, $H_I = 10^{13}\,\mathrm{GeV}$, $N_* = 60$, and $|\Phi_i| = 10^4 F_a N$. The constraint $|\Phi_i| < M_{\rm p}$ corresponds to the region below the red dotted line.
}
\label{Region}
\end{figure}

The axion parameter space in $(F_a, N)$ that satisfies \geqn{eq1} and \geqn{eq2}, obtained by numerical solving, is shown in \gfig{Region}. The isocurvature constraint requires a larger $|\Phi_i|$, or equivalently a larger $F_a$, so that the axion obtains a larger effective mass and its fluctuations are suppressed. On the other hand, the axion quality constraint requires a larger $N$, since this pushes the leading Planck-suppressed PQ-breaking operator to higher order. Therefore, the allowed region is concentrated in the upper-right corner of the figure. However, neither of these quantities can be increased without bound, because they are jointly constrained by $|\Phi_i| < M_{\rm p}$.

This observation is also phenomenologically interesting. If one believes that high-scale inflation and the high quality of the axion should both be guaranteed by a discrete gauge symmetry with a large domain wall number, then our scenario leads to a prediction of axion decay constant lying in the range
$F_a \sim 5 \times 10^{12} - 10^{13}\,\mathrm{GeV}$. Using
$m_a \simeq 5.7\,\mu\mathrm{eV}\,(10^{12}\,\mathrm{GeV}/F_a)$, this corresponds to
\begin{equation}
    m_a \simeq 0.6 - 1.2\,\mu\mathrm{eV}.
\end{equation}
This mass range can be probed by high-precision terrestrial axion haloscope experiments~\cite{Adams:2022pbo,Rybka:2024axion,Ma:2026ujz}.
Second, high scale inflation with $H_I \sim 10^{13}\,$GeV the predicts a tensor-to-scalar ratio of order $r \simeq 10^{-3}$.
This is below the current observational bound, but may be accessible to future tensor mode observations with improved sensitivity~\cite{LiteBIRD:2022cnt,Belkner:2023bmh}. 

A simple ultraviolet realization of the large discrete symmetry considered above can be obtained in a KSVZ-type construction with multiple heavy quark pairs~\cite{Kim:1979if,Shifman:1979if}. One introduces the complex PQ field $\Phi$ and $N_Q$ pairs of vector-like colored fermions $Q_i,  \bar{Q}_i$, which are vector-like under the Standard Model gauge group but chiral under the PQ symmetry. The Yukawa interactions $y_i \Phi Q_i \bar Q_i$ generate heavy-quark masses after $\Phi$ develops a vacuum expectation value. Since the QCD anomaly coefficient receives an additive contribution from each heavy quark pair, choosing $N_Q$ fundamental pairs with the same PQ charge assignment gives $N_{\rm DW} = N_Q$. In the minimal realization relevant for our discussion, one may take $Z_N$ symmetry with $N = N_Q = N_{\rm DW}$, so that the same large integer controls both the axion quality and the domain wall number. Such KSVZ constructions with discrete gauge symmetries and with multiple heavy quarks have been studied in the literature~\cite{Plakkot:2021hqe,Dias:2002gg}.
A possible KSVZ realization with many vector-like heavy quark pairs is also subject to a perturbativity constraint from the QCD running. 
For $N$ fundamental heavy quark pairs with masses around $M_Q\simeq 10^{14}\,{\rm GeV}$, the one-loop beta-function coefficient above the threshold becomes $b_0=7-2N/3$, so QCD loses asymptotic freedom for $N\geq 11$ and the coupling grows toward the planck scale. 
Requiring the strong coupling to remain perturbative up to the reduced Planck scale gives roughly $N\lesssim 35$ \footnote{If the $Q_i$ and the $\bar Q_i$ form a family and an anti-family we obtain maximally 7 pairs of the family and the anti-family with one family-antifamily pair contributes $N_{\rm DW}=4$. The constraint on the number of the pairs is coming from the perturbativity condition of the $U(1)_Y$ gauge interactions}.

One further comment is that, to avoid parametric resonance, here we require the ratio between the early-time and late-time PQ field values to be bounded roughly as
$10^4$. 
However, if the PQ potential contains an additional sextic term, as discussed in \gsec{sec:parameter}, the early-time and late-time field values can be partially decoupled. In that case, even if $|\Phi_i| \sim M_{\rm p}$ is required during inflation in order to generate a sufficiently large effective axion mass and suppress isocurvature fluctuations, the final value of $v_{\rm PQ}$ can still remain comparatively small. This opens up the possibility that the same mechanism can work with a smaller discrete symmetry number $N$, since the axion quality constraint becomes less severe for a smaller final vev.


\section{Conclusion and Discussions}

The QCD axion is a well-motivated DM candidate, since it can simultaneously account for the observed DM abundance and solve the strong CP problem. However, axion DM faces serious theoretical and cosmological challenges. On the theoretical side, the PQ symmetry must be of very high quality: Planck-suppressed operators that explicitly violate the global $U(1)_{\rm PQ}$ symmetry can shift the axion minimum and spoil the solution to the strong CP problem. On the cosmological side, if the PQ symmetry is broken during high-scale inflation, the axion generically acquires quantum fluctuations, which lead to cold DM isocurvature perturbations exceeding the CMB bound.

A standard way to address the axion quality problem is to impose a discrete gauge symmetry, which forbids dangerous low-dimensional PQ-violating operators and pushes the leading explicit breaking term to sufficiently high dimensional operators. However, such constructions often require a large discrete symmetry, which is identified with a large axion domain wall number. This makes the usual Linde mechanism, which assume a large PQ field value during inflation, less effective, because the axion fluctuation is enhanced by the domain wall number even when the PQ field takes a large value during inflation. 

In this work, we have proposed a new mechanism in which the high quality of axion protected by a discrete gauge symmetry $Z_N$ naturally avoids the isocurvature problem. The key point is that the leading operator allowed by the discrete symmetry, while violating the PQ symmetry, induces a large effective axion mass during inflation when the PQ field has a large field value. As a result, the axion fluctuation is exponentially suppressed. In this sense, the same discrete symmetry that protects the axion quality from quantum gravity effects provides a natural solution to the axion isocurvature problem.

By imposing both the axion quality bound and the isocurvature constraint, we found that, for the benchmark parameters considered in this work, the axion decay constant should be in the range
$F_a \sim 5 \times 10^{12} - 10^{13}\,{\rm GeV}$, corresponding to an axion mass
$m_a \sim 0.6 - 1.2\,\mu{\rm eV}$. It can be tested by future axion haloscope searches.
However, our mechanism works only when the 
discrete symmetry or domain wall number is about $N_{\rm DW} = 18 \sim 35$. 

\section*{Acknowledgements}

M. K. is supported by JSPS KAKENHI Grant No. 25K07297. J. S. is supported by the Japan Society for the Promotion of Science (JSPS) as a part of the JSPS Postdoctoral Program (Standard) with grant number: P25018. T. T. Y. is supported by the Natural Science Foundation of China (NSFC) under Grant No. 12175134, MEXT KAKENHI Grants No. 24H02244.
All authors are supported
by the World Premier International Research Center Initiative (WPI), MEXT, Japan (Kavli IPMU).

\providecommand{\href}[2]{#2}\begingroup\raggedright\endgroup

\end{document}